# Perturbative predictions for $B_c$ meson production in hadronic collisions

Marco Masetti

Dipartimento di Fisica, Università di Roma "Tor Vergata",

and I.N.F.N., Sezione di Roma II, Viale della Ricerca Scientifica

I-00133 Roma, Italy

Francesca Sartogo

Dipartimento di Fisica, Università di Roma "la Sapienza",

and I.N.F.N., Sezione di Roma, Piazzale A. Moro 2,

I-00185 Roma, Italy

March 31, 1995

**Abstract**

Perturbative cross section for direct $B_c$ meson production in gluon gluon scattering $gg \to B_c^+ b\bar{c}$ is calculated and compared with other existing results. Predictions for hadronic $B_c$ production at Tevatron and LHC are presented and the main sources of uncertainties are discussed.

In the last years many theoretical studies on $B_c$ meson properties were published, but an experimental identification of this pseudoscalar bound state formed by two quarks with different flavours is still missing, due to the very small production cross section.

The mass spectrum of $\bar{b}c$ bound states is expected to be similar to the spectra of other mesons with two heavy quarks (charmonium and bottomonium) and predictions based on potential models and on QCD sum rules [1, 2] indicate the presence of a rich structure of excited states below the fragmentation threshold in a $B$ and a $D$ meson. There is however an important difference with respect to unflavoured quarkonia: the excited states cannot decay directly into light mesons, but decay only to the pseudoscalar $B_c$ ground state, due to the flavour conservation in strong and electromagnetic interactions. The lifetime of the weak-decaying $B_c$ meson is expected to be in the range of $D$ mesons lifetimes [3, 4]. Exclusive $B_c$ decays were studied by several authors [3, 4, 5]. Some decay modes are interesting for experimental detection, for example the semileptonic decay with three charged leptons in the final state ($B_c \to J/\psi + l + \nu_l$ followed by the electromagnetic decay $J/\psi \to l'^+ + l'^-$) with composite branching ratio of about $10^{-2}$, and the decay in $J/\psi + \pi$, with branching ratio of the order of $10^{-3}$. The lack of experimental detection of $B_c$ mesons is a consequence of the small production cross section: in fact the production via electromagnetic or strong interaction requires two additional heavy quarks in the final state, and is therefore suppressed at typical experimental energies. The production of a $B_c$ without other heavy particles should be possible in processes like $e + \nu \to W^* \to B_c$, which however involve weak interactions and are strongly suppressed due to the high virtuality of the $W^*$.

The standard estimates of $B_c$ production cross section are based on perturbative calculations with the heavy quark bound state described in the zero binding energy limit. The simplest but important case of $B_c$ production in $e^+ e^-$ annihilation involves only four Feynman diagrams and was studied in detail [6]. The total branching ratio for $Z^0$ with a $B_c^+$ or a $B_c^-$ in the final state (including the contribution due to the production of excited $B_c$ states which subsequently decay to the ground state) is predicted to be $140 \cdot 10^{-6}$ i. e. $\sigma(Z^0 \to B_c^\pm X) \sim 10^{-3} \sigma(Z^0 \to b\bar{b} X)$. An estimate obtained with the Montecarlo event generator Herwig, which for $Z^0$ decays in $B_c^\pm$ gives similar results, indicates that also at high energy hadronic colliders like Tevatron and LHC the $B_c^\pm$ production cross section would be around $10^{-3}$ times the $b\bar{b}$ production cross section [7].

In the perturbative approach the relevant process for $B_c$ hadroproduction is the gluon gluon scattering $g\, g \to B_c^+\, b\, \bar{c}$, which at the lowest order involves 36 diagrams. The process $q\bar{q} \to B_c^+\, b\, \bar{c}$ is expected to give a negligible contribution to the hadronic $B_c$ production cross section. The calculation is extremely simplified if one considers only the *fragmentation* terms [8], which were expected to dominate at high energy; however this approach fails at least in the photoproduction process, as shown in [9]. The complete perturbative calculation for $g\, g \to B_c^+\, b\, \bar{c}$ was performed by several authors [11, 12, 13], with results in disagreement with each other.

From this summary of theoretical predictions about the $B_c$ mesons it should be clear that high energy hadronic colliders give the best chances to identify these particles, and it is therefore necessary to reduce the large uncertainties due to the discrepancies between existing predictions. For this reason in this letter we present the results of an independent study on direct $B_c$ hadroproduction. The dependence of the results on different choices of the gluon distribution functions and on the scale for $\alpha_s$, which are the main sources of uncertainty, is discussed. Our results are definitely incompatible with [11] and [13], which however give predictions differing with each other at least by a factor 20, difficult to ascribe only to different choices of gluon distribution functions. On the other hand our results are essentially in agreement with [12].

The calculation technique is standard: in the approximation of negligible binding energy and relative momentum in the bound state, which is justified for bound states with only heavy constituents, the quarks are on shell and have parallel momenta ($m_b p_c^\mu = m_c p_b^\mu$) and the meson mass is $M_{B_c} = m_c + m_b$. The 36 Feynman diagrams for the process $g\, g \to B_c^+\, b\, \bar{c}$ can be obtained from the 13 diagrams of fig. 1 performing all possible interchanges of initial gluon momenta and of final quark flavours. The rule to describe the bound state in this approximation is to insert as 'bound state vertex' the spin projector

$$\frac{\delta^{cc'}}{\sqrt{N_c}} \frac{f_{B_c}}{\sqrt{48}} \Gamma\, (\not{p} + M_{B_c})$$

where $c$ and $c'$ are colour indices and $\Gamma = \gamma_5$ for a pseudoscalar meson while $\Gamma = \not{\epsilon}$ for a vector meson. The decay constant $f_{B_c}$ is defined by

$$<0|\bar{b}\gamma_\mu\gamma_5 c|B_c(P)> = if_{B_c} P_\mu$$

and in the non relativistic limit it is proportional to the wave function in the origin:

$$f_{B_c} = \sqrt{\frac{12}{M_{B_c}}}|\Psi(0)|.$$

The simple substitution of the sum on external gluon polarizations $\sum_{i=1,2} \varepsilon_{i\mu}\varepsilon_{i\nu}$ with $g_{\mu\nu}$ would give ghost contributions. It is possible to avoid ghosts summing only on physical transversal polarizations:

$$\sum_{i=1,2} \varepsilon_{i\mu}\varepsilon_{i\nu} = -g_{\mu\nu} + \frac{k_\mu n_\nu + k_\nu n_\mu}{k \cdot n} - \frac{k_\mu k_\nu}{(k \cdot n)^2}$$

where $k$ is the gluon momentum and we choose $n^\mu = (k_a{}^\mu + k_b{}^\mu)/\sqrt{\hat{s}}$. The number of independent terms in the cross section can be greatly reduced using relations between amplitudes connected by exchange of the initial gluons or of final quarks flavours and momenta. Using the algebraic program Schoonschip [14] we calculate the squared amplitude and integrate analytically to obtain the differential partonic cross section $d^2\hat{\sigma}/dp_T dy$, where $p_T$ is the transverse momentum and $y$ is the c.m. rapidity of the produced $B_c$. The subsequent integrations are performed both using the Montecarlo integration program Vegas [15] and using simple algorithms for numerical evaluation of low dimensional integral, to check the convergence of the result.

| $\sqrt{\hat{s}}$(GeV) | $\hat{\sigma}$ | $\hat{\sigma}$ [11] | $\hat{\sigma}$ [12] |
|---|---|---|---|
| 20 | 23.6 | 5.3 | 22.5 |
| 30 | 26.0 | 9.2 | — |
| 40 | 22.3 | — | 30.8 |
| 60 | 15.5 | 8.6 | — |
| 80 | 11.2 | — | — |
| 100 | 8.4 | — | 16.2 |

Table 1: Cross section (in pb) for $g\,g \to B_c^+ \, b \, \bar{c}$. All the results are rescaled to $f_{B_c} = 500$ MeV.

As a preliminary step and a check of our calculation we studied the direct production in photon photon collisions $\gamma\,\gamma \to B_c^+ \, b \, \bar{c}$. At the lowest order of perturbation theory this process involves 20 diagrams which can be obtained from those of fig. 1 eliminating the diagrams containing non abelian vertices, and replacing the external gluons with photons. This calculation was already performed in [9] and [10]. Our results for this process, using the same values for the parameters $m_b$, $m_c$, $f_{B_c}$, $\alpha_s$, $\alpha_{em}$, are in good agreement with those of [9] and [10].

For the process $g\,g \to B_c^+ \, b \, \bar{c}$ we use the following values of the parameters:

$$f_{B_c} = 500 \text{ MeV}, \quad \alpha_s = 0.2, \quad M_{B_c} = 6.3 \text{ GeV}, \quad m_b/m_c = 3$$

Note that the partonic cross section is a function of the kinematical variables times $f_{B_c}^2 \alpha_s^4$, and therefore the rescaling of the results for different values of $f_{B_c}$ and $\alpha_s$ is easy to obtain.

The total cross section for this process as a function of the energy in the partonic center of mass $\sqrt{\hat{s}}$ is plotted in fig. 2, while the differential partonic cross sections $d\hat{\sigma}/dp_T$ and $d\hat{\sigma}/dy$ are shown for different values of $\hat{s}$ in fig. 3 and fig. 4.

In Tab. 1 the total partonic cross section for several values of $\hat{s}$ is compared with the analogous existing calculations [11, 12], rescaling all the results to the same value of $f_{B_c}$. Unfortunately the comparison is made difficult by the small amount of information on the partonic cross section before the convolution with the gluon distribution functions. The results of [11] are significantly smaller than ours. Even if it is not clear to us if their results refer to a fixed value of $\alpha_s$ or a running one, and in this last case which scale they choose, the difference cannot be explained simply by the running of $\alpha_s$.

In [12] $\alpha_s$ is fixed to 0.2 and $f_{B_c} = 570$ MeV; with the appropriate rescaling of $f_{B_c}$ our results are in reasonable agreement with theirs in the low $\hat{s}$ region. Given the behaviour of the partonic cross section, the discrepancies between the values we obtain and those obtained by [12] at higher energies are not really relevant in the evaluation of the hadronic cross section, after the convolution with the gluon distribution functions. Finally, a comparison with [13] is impossible at this level, since only results on the hadronic cross section are discussed there.

| $\sqrt{s}$(TeV) | | gluon distribution | scale | $\sigma$ (nb) | $\sigma$ (nb) *this paper* |
|---|---|---|---|---|---|
| 0.12 | [11] | EHLQ | $\hat{s}/4$ | 4.5 $10^{-3}$ | 60. $10^{-3}$ |
| 1.8 | | | | 1.0 | 8.5 |
| 16. | | | | 12.3 | 78 |
| 1.8 | [12] | [12] | $\hat{s}$ * | 9.4 | 8.8 |
| 16. | | | $\hat{s}$ * | 151 | 140 |
| 1.8 | [13] | CTEQ | $\hat{s}$ | 18.3 | 2.5 |
| 1.8 | | | $\hat{s}/4$ | 29.6 | 3.8 |
| 1.8 | | | $4M_{B_c}^2$ | 31.5 | 3.9 |

Table 2: Comparison of hadronic cross section (in nb). All results are rescaled to $f_{B_c} = 500$ MeV. In the fourth column we indicate the scale for gluon distribution functions and running $\alpha_s$; the * indicates results obtained for fixed $\alpha_s = 0.2$, evolving only the gluon distribution. In the fifth and sixth column the cross sections obtained in the cited papers and the values we obtain using the same prescriptions are reported.

The cross section for the hadronic processes $p\,p \to B_c^+ X$ and $p\,\bar{p} \to B_c^+ X$ can be obtained by the convolution of the partonic cross section with the gluon distribution functions $g(x, Q^2)$:

$$\sigma(s) = \int_0^1 dx_1 \int_0^1 dx_2 g(x_1, Q^2) g(x_2, Q^2) \theta(x_1 x_2 s - 4M_{B_c}^2) \hat{\sigma}(\hat{s} = x_1 x_2 s).$$

In figs. 5 and 6 the differential cross section in $p_T$ and $y_{lab}$ obtained with the MRS(A) gluon distribution function [16] is plotted for the energies corresponding to Tevatron (1.8 TeV) and LHC (14 TeV).

In Tab. 2 we compare the values for direct $B_c^+$ production cross section in proton proton collisions reported in [11, 12, 13], with the results we obtain using the same prescriptions and gluon distribution functions (EHLQ [18], CTEQ [19] and the parametrization described in [12]). In [11] the EHLQ gluon distribution function is used fixing the evolution scale $Q^2$ at $\hat{s}/4$. With the same gluon distribution and the same $Q^2$ we obtain for the hadronic cross section values greater of those of [11] especially at the lower energies. This comparison confirms the disagreement with [11] already observed for the partonic cross section.

To compare our predictions with [12], we fix $\alpha_s = 0.2$ as they do, and we use the same parametrization of gluon distribution function they describe in [12] with an evolution scale $Q^2 = \hat{s}$. There is agreement between our results and those of [12], especially at the lower energy. The results of [13] are much larger than all other existing predictions. We obtain results which are smaller than those of [13] by at least a factor 7, using the same gluon distribution (CTEQ) and the same evolution scale as they do, and therefore we conclude that the incompatibility of our results with those of [13] is not related to the choice of gluon distribution functions and can be only a consequence of a disagreement in the calculation of the partonic cross section $\sigma(g\,g \to B_c^+ \,b\,\bar{c})$.

| gluon distribution | $Q^2$ | $\sigma$ (nb) $\sqrt{s} = 1.8$ TeV | $\sigma_0$ (nb) $\sqrt{s} = 1.8$ TeV | $\sigma$ (nb) $\sqrt{s} = 14$ TeV | $\sigma_0$ (nb) $\sqrt{s} = 14$ TeV |
|---|---|---|---|---|---|
| **MRS(A)** | $\hat{s}$ | **3.2** | 7.3 | **44** | 110 |
| MRS(A) | $4M_{B_c}^2$ | 4.9 | 6.6 | 61 | 80 |
| MRS(G) | $\hat{s}$ | 3.3 | 7.4 | 57 | 139 |
| MRS(G) | $4M_{B_c}^2$ | 5.2 | 6.9 | 82 | 109 |
| GRV | $\hat{s}$ | 3.9 | 8.8 | 81 | 196 |
| **GRV** | $4M_{B_c}^2$ | **6.4** | 8.6 | **121** | 163 |
| [12] | $\hat{s}$ | 3.8 | 8.8 | 47 | 121 |
| [12] | $4M_{B_c}^2$ | 5.8 | 7.8 | 62 | 84 |
| EHLQ | $\hat{s}$ | 5.6 | 9.4 | 60 | 113 |
| EHLQ | $4M_{B_c}^2$ | 8.8 | 8.1 | 77 | 71 |

Table 3: Hadronic cross section (in nb) with $f_{B_c} = 500$ MeV; $Q^2$ is the scale for gluon distribution functions; $\sigma$ and $\sigma_0$ are the cross sections calculated with running $\alpha_s(Q^2)$ and with fixed $\alpha_s = 0.2$ respectively. We put in evidence the maximum and minimum value obtained for running $\alpha_s(Q^2)$ using the more updated gluon distribution parametrizations.

In Tab. 3 the results we obtain with several gluon distribution functions (MRS(A) and MRS(G) [16], GRV [17], EHLQ [18], and the parametrization of [12]) are shown. While the choice of the gluon distribution function parametrization EHLQ and [12] is mainly dictated by the comparison with other existing results, it is also interesting to show the values obtained with other updated sets of gluon distribution functions. In order to discuss the possible dependence of the results on the evolution scales for the gluon distributions and for the running coupling constant $\alpha_s$, we vary $Q^2$ between $4M_{B_c}^2$ and $\hat{s}$. The case of fixed $\alpha_s$ ($\alpha_s = 0.2$) is also considered, as suggested in [12]. We will not attempt to justify a particular choice of the evolution scale in the leading order calculation described here; it is however clear that results obtained fixing $\alpha_s = 0.2$ must be considered as optimistic higher limits on the predicted cross sections. It should be noted that the relatively high result one obtains using EHLQ and $\alpha_s$ running is due to the relatively high value of $\Lambda_{QCD}$ rather than to the shape of the gluon distribution function. The difference in the choice of the gluon distribution function however emerges more clearly at the energy corresponding to LHC. While at the energy of Tevatron the relevant $x$ region is $x \gtrsim 10^{-4}$, in the case of LHC it becomes $x \gtrsim 10^{-6}$. All gluon distribution parametrizations are expected to be valid only above a given $x_{min}$ (a typical value is $x_{min} = 10^{-5}$) and therefore the cross sections at higher energies are typically underestimated. The only exception is the GRV gluon distribution function, for which $x_{min} = 10^{-6}$, but it has to be noted that the shape of the GRV gluon is considerably steeper than, for example, the MRS(G) gluon, which comes from a recent analysis of the latest measurements of the structure function $F_2$ at HERA [16]. The predicted cross section for direct $B_c$ production at Tevatron and LHC, taking into account the uncertainties on the knowledge of the behaviour of the gluon at very small $x$, can be obtained with MRS and GRV gluon distributions:

$$\sigma_{direct}(p\bar{p} \to B_c^+ X) = \frac{f_{B_c}^2}{(500\ MeV)^2}\ (3.2\ -\ 6.4)\ \text{nb} \qquad \text{(Tevatron; 1800 GeV)};$$

$$\sigma_{direct}(pp \to B_c^+ X) = \frac{f_{B_c}^2}{(500\ MeV)^2}\ (0.04\ -\ 0.12)\ \mu\text{b} \qquad \text{(LHC; 14 TeV)}.$$

A possible important source of uncertainty in the predicted cross sections is the factor $f_{B_c}^2$. The existing predictions vary from $f_{B_c} = 160$ MeV [20] to $f_{B_c} = 600$ MeV [21]. Varying $f_{B_c}$ between so different values would give a contribution to the uncertainty much larger than those due to the choices of parton distributions and of evolution scales. A more careful inspection of the predictions shows that potential models have a much smaller uncertainty ($f_{B_c} = 500 \pm 80$ MeV: for a detailed discussion and review of predictions for $f_{B_c}$ see [2]), while QCD sum rules give results spread in a wide interval, due to the intrinsic ambiguities of the method. However in [2] it is asserted that the most reliable predictions from QCD sum rules are compatible with potential models ($f_{B_c} = 460 \pm 60$ MeV). The simpler way to obtain predictions for $f_{B_c}$ and for the $c\bar{b}$ spectrum is to rescale the charmonium and bottomonium spectra to the intermediate case of $c\bar{b}$ bound states: this is the role of potential models which can be considered as a sort of parametrization of quarkonium spectra which gives predictions for $B_c$. For this reason it seems to us that potential models are presently more reliable than QCD sum rules for phenomenological predictions on $c\bar{b}$ spectrum and wave functions. Considering only potential model predictions, the uncertainty in $B_c$ production cross sections due to the factor $f_{B_c}^2$ is about 30%, much smaller than the uncertainty from $\alpha_s$ and gluon distributions.

Finally we want to give an estimate of the total $B_c$ production cross section in hadronic collisions, including the contribution due to the production of excited $c\bar{b}$ states which subsequently decay to pseudoscalar $B_c$. The main contribution should come from vector $B_c^*$ production. In [12] this cross section is calculated obtaining $\sigma(pp \to B_c^* X) \simeq 2.5\sigma(pp \to B_c X)$. Neglecting further contributions from other excited $B_c$ states we conclude that the total cross section for the production of $B_c^+$ and $B_c^-$ would be

$$\sigma_{total}(p\bar{p} \to B_c^+ X) \sim 20\ -\ 50\,\text{nb} \qquad \text{(Tevatron; 1800 GeV)};$$

$$\sigma_{total}(pp \to B_c^+ X) \sim 0.3\ -\ 0.8\,\mu\text{b} \qquad \text{(LHC; 14 TeV)}.$$

This correspond to at least one thousandth of the total $b\bar{b}$ production cross section.

In this letter we discussed the main sources of uncertainties in the calculation of the cross section for direct $B_c$ meson production in hadronic collisions at the lowest order of perturbation theory and in the zero binding energy limit. The uncertainties related to the choice of input parameters and of gluon distribution parametrizations amount roughly to a factor 3, too small to explain the large discrepancies between the different existing calculations. Finally we gave an estimate of the total $B_c$ production cross section, including the contribution of excited states, for Tevatron and LHC.

We thank M. Lusignoli for having drawn our attention to the problem treated in this letter and for many useful discussions and comments. We are also grateful to A. K. Likhoded, V. Lubicz, S. Petrarca and B. Taglienti.

# Figure Captions

[1] Feynman diagrams for $g\ g \to B_c^+\ b\ \bar{c}$. The complete set of 36 diagram can be obtained from the diagrams shown in the figure performing all possible interchanges of initial gluon momenta and of final quark flavours.

[2] Total partonic cross section $\hat{\sigma}(g\ g \to B_c^+\ b\ \bar{c})$ (in pb).

[3] Partonic $p_T$ distribution $d\hat{\sigma}/dp_T$ (in pb GeV$^{-1}$) for $\sqrt{\hat{s}} = 20$, 30, 40, 60, 80 and 100 GeV (in order of increasing $p_T$ endpoint).

[4] Partonic rapidity distribution $d\hat{\sigma}/dy$ (in pb) for $\sqrt{\hat{s}} = 20$, 30, 40, 60, 80 and 100 GeV (in order of increasing rapidity endpoint).

[5] Differential cross section $d\sigma/dp_T$ (in nb GeV$^{-1}$) for direct $B_c$ production calculated with MRS(A) gluon distribution function for LHC ($\sqrt{s} = 14$ TeV) with the evolution scale $Q^2 = 4M_{B_c}^2$ (upper solid line) and $Q^2 = \hat{s}$ (lower solid line), and for Tevatron ($\sqrt{s} = 1.8$ TeV) with the evolution scale $Q^2 = 4M_{B_c}^2$ (upper dotted line) and $Q^2 = \hat{s}$ (lower dotted line).

[6] Differential cross section $d\sigma/dy_{lab}$ (in nb) for direct $B_c$ production calculated with MRS(A) gluon distribution function for LHC ($\sqrt{s} = 14$ TeV) with the evolution scale $Q^2 = 4M_{B_c}^2$ (upper solid line) and $Q^2 = \hat{s}$ (lower solid line), and for Tevatron ($\sqrt{s} = 1.8$ TeV) with the evolution scale $Q^2 = 4M_{B_c}^2$ (upper dotted line) and $Q^2 = \hat{s}$ (lower dotted line).

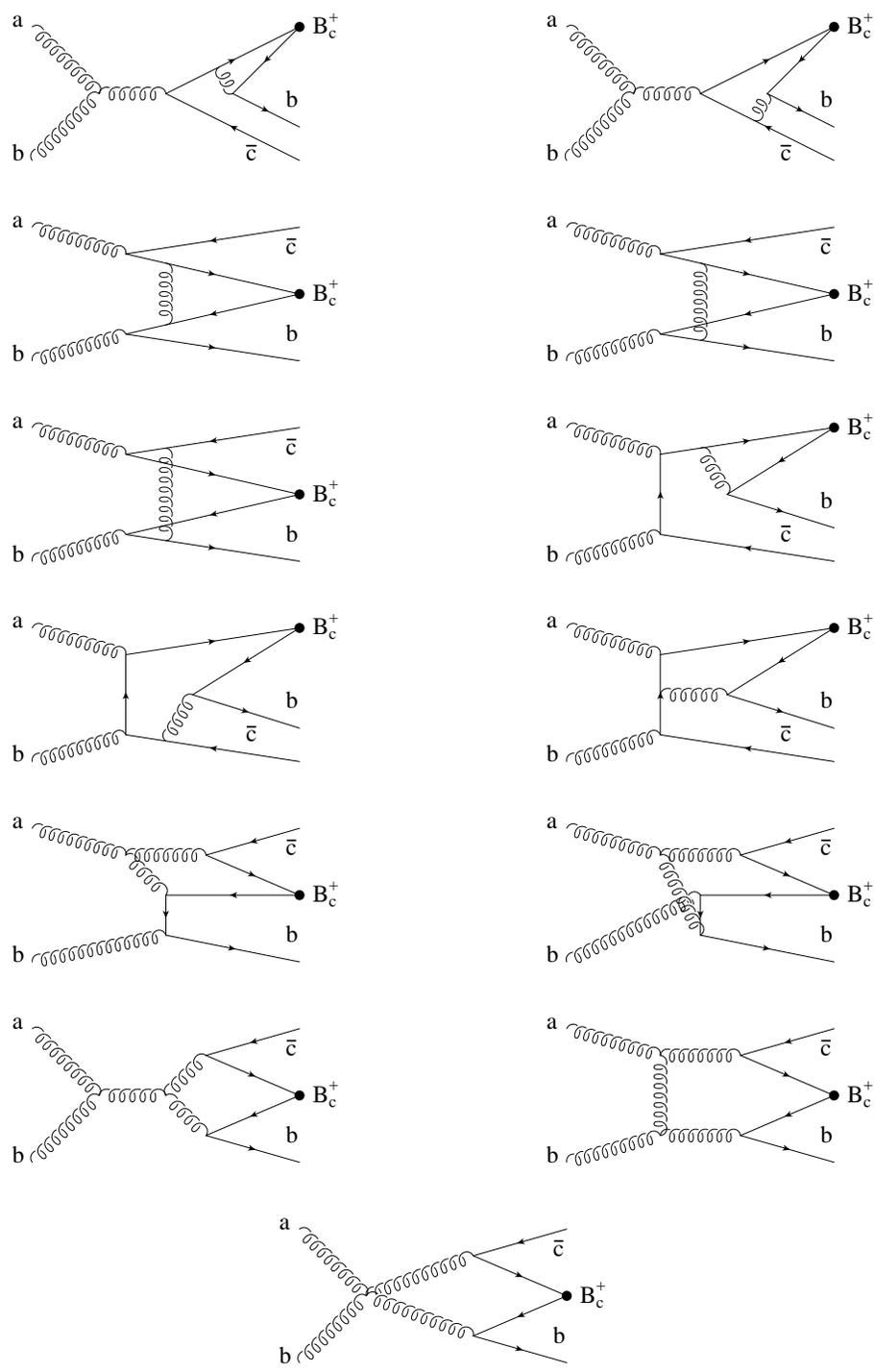

Figure 1:

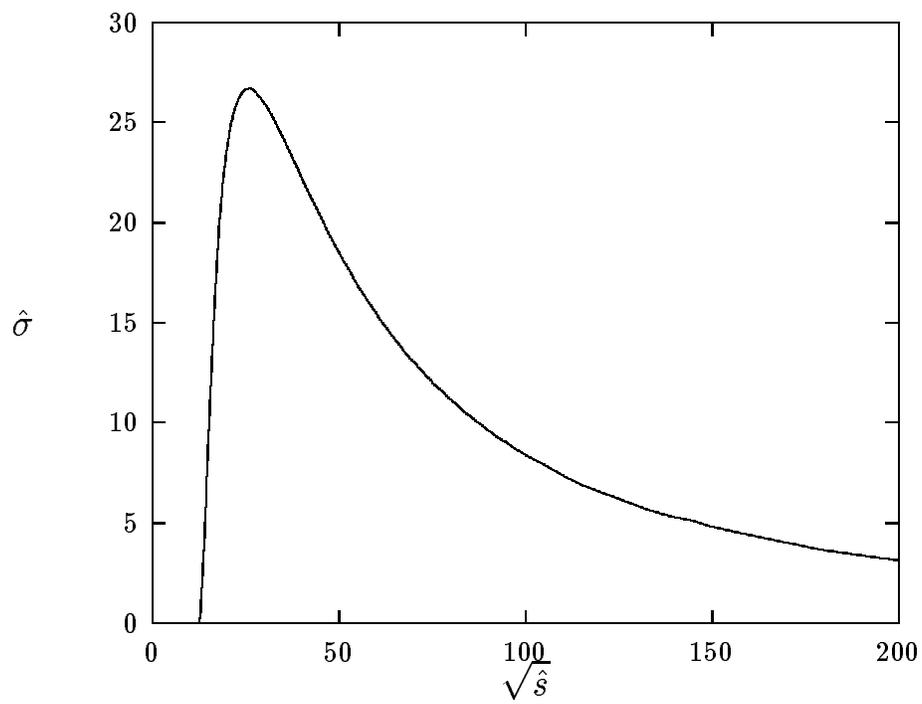

Figure 2:

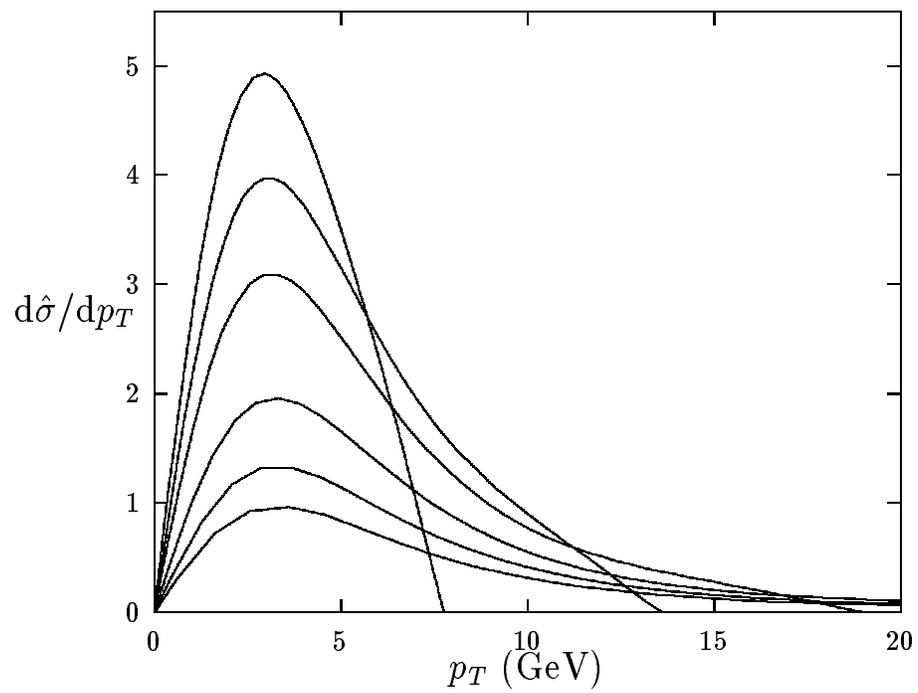

Figure 3:

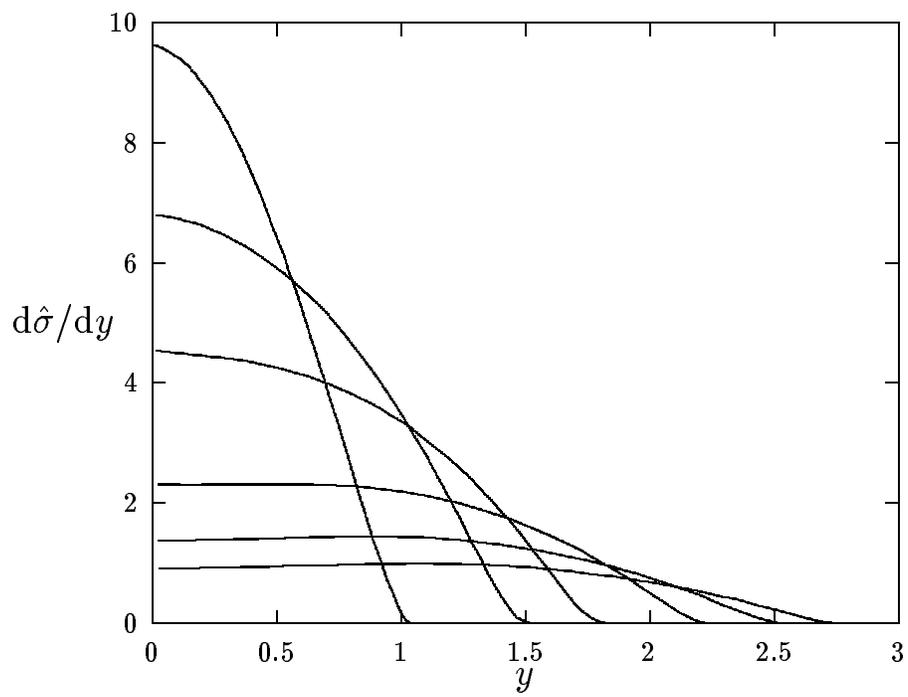

Figure 4:

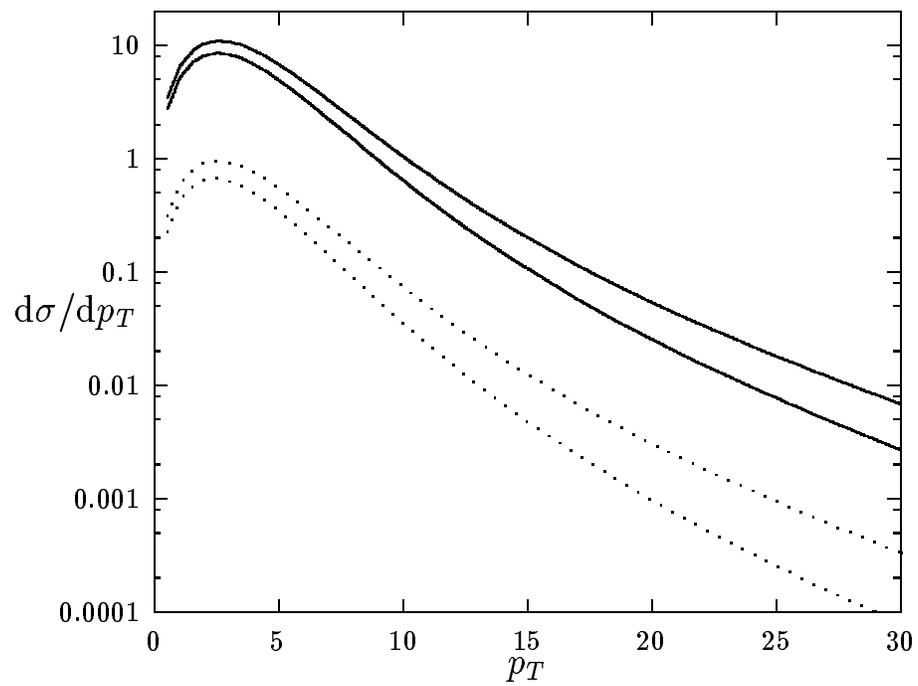

Figure 5:

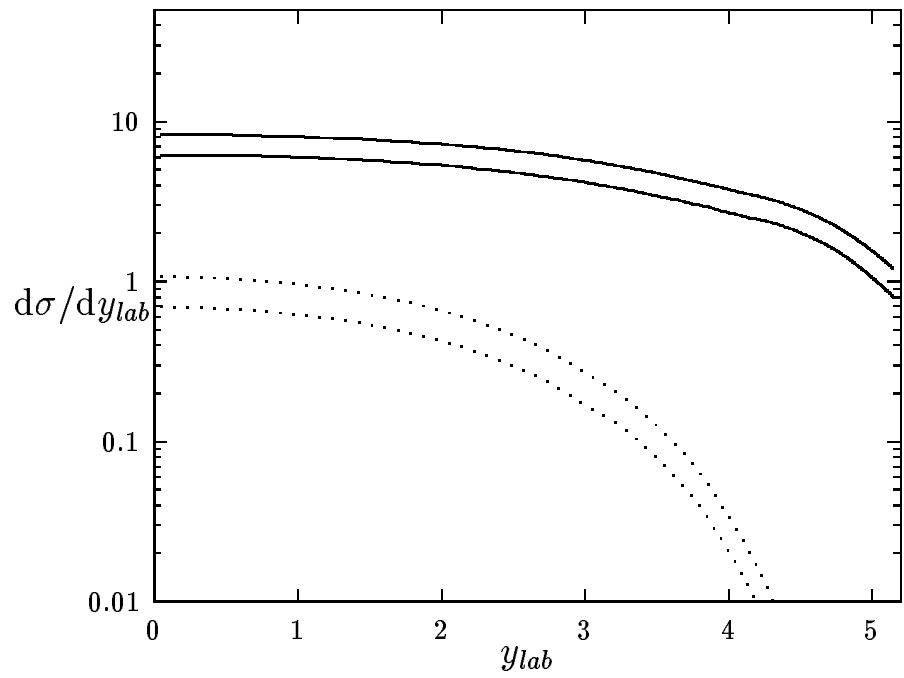

Figure 6: